\documentclass{article}

\usepackage{a4}
\usepackage{graphicx}

\begin{document}


\title{PENETRATING  RADIATION AT THE SURFACE OF AND IN WATER}
\author{Note by D. PACINI; Nuovo Cimento VI/3 93 (1912)}
\date{Translated and commented by Alessandro De Angelis
\\INFN and University of Udine}


\maketitle

\noindent {\bf{Foreword -- }}
At the beginning of the twentieth century, two scientists, the
Austrian Victor Hess\footnote{V.F. Hess, Phys. Zeit. 13, 1084-1091, November 1912.} and the Italian Domenico Pacini\footnote{D. Pacini, Nuovo Cimento VI/3, 93-100, February 1912.},
developed two brilliant lines of research independently, leading to the
determination of the origin of atmospheric radiation. Before their
work, the origin of the radiation -- today called ``cosmic rays'' -- was
strongly debated, as many scientists thought that these particles
came from the crust of the Earth.

The approach by Hess is well known: Hess measured the rate of
discharge of an electroscope that flew aboard an atmospheric balloon.
Because the discharge rate increased as the balloon flew at higher
altitude, he concluded in August 1912 that the origin could not be
terrestrial. For this discovery, Hess was awarded the Nobel Prize in
1936, and his experiment became legendary.

Shortly before, in June 1911,  Pacini, a professor at the University of Bari,
made a series of measurements to determine the variation in
the speed of discharge of an electroscope (and thus the intensity of
the radiation)  while the electroscope was immersed in a box in the sea
near the Naval Academy in the Livorno  Bay (the Italian Navy
supported the research). The measures are
documented in his work$^2$ \emph{La radiazione penetrante alla superficie ed in
seno alle acque} \emph{(Penetrating radiation at the surface of and in 
water).}  Pacini discovered (italics are in the original) that the discharge of the oscilloscope was significantly slower
than at the surface: 
\begin{quote}{``The apparatus [...]  was enclosed in a copper box to be
able to immerse in
 depth.  [...] The experiments were performed [...] with the apparatus on the surface   and immersed at a depth of 3 meters. 
[...It] appears from the results of the work described in this Note: that \emph{a sizable cause of ionization exists in the atmosphere, originating from penetrating radiation, independent of the direct action of radioactive substances in the soil.}"}\end{quote}

Documents testify that Pacini and Hess knew of each 
other's work\footnote{{P. Carlson and A. De Angelis,}  {\em Nationalism and internationalism in science: the case of the discovery of cosmic rays,}  Eur. Phys. J. H, in press (arXiv:1012.5068 [physics.hist-ph]).}$^,$\footnote{{A. De Angelis,}  {\em Domenico Pacini, uncredited pioneer of the discovery of cosmic rays,} Rivista del Nuovo Cimento 33, 713-756, 2010.}. Pacini died in 1934, two years before the Nobel Prize was awarded for
the discovery of cosmic rays.
While Hess is remembered as the
discoverer of  cosmic rays, the simultaneous discovery by Pacini is
forgotten by most.

\vskip 2mm

\begin{flushright} Alessandro De Angelis \end{flushright}


\newpage



\begin{center}
{\center{\bf{PENETRATING  RADIATION AT THE SURFACE OF\\AND IN WATER}}}

{\center{Note by D. PACINI}}

{\center{Translated and commented by Alessandro De Angelis\\INFN and University of Udine}}
\end{center}

\vskip 1cm

Observations that were made on the sea during the year 1910\footnote{D. Pacini. Ann. dell'Uff. Centr. Meteor. Vol. XXXII, parte I, 1910.\\- Le Radium, T. VIII, pag. 307, 1911.} led me
to conclude that a significant proportion of the pervasive radiation
that is found in air had an origin
that was independent of direct action of the active substances
in the upper layers of the Earth's surface.

Here, I will report on further experiments that support
this conclusion.

The results that were previously  obtained indicated that a source of ionization existed on
the sea surface, where  possible effects from the soil are small, that had such an intensity that 
could not
be explained on the basis of the known distribution
of radioactive substances in water and in air.

Indeed, as shown by Eve\footnote{A.S. Eve. Phil. Mag., 1911.}, one can easily calculate
 the expected ionizing action at the surface of the 
sea, 
due to
$\gamma$ radiation that is emitted by active particles in air.

Let:
\begin{itemize}
\item[-] $Q$  be the equivalent radiation in Ra. C per cm$^3$ in the atmosphere, expressed
in grams of Radium in radioactive equilibrium, $Q = 8 \times 10^{-17}$.
\item[-]  $K$ be the number of ions that is generated per cm$^3$ per second from one gram
of Radium at a distance of 1~cm:  $K = 3.4 \times 10^9$ for air enclosed
in an Aluminum electroscope;  $K = 3.1 \times 10^9$ for free air.
\item[-]  $\lambda$ be the absorption coefficient of $\gamma$ rays in air =
0.000044.
\item[-]  $r$ be the distance from the point at which we consider the action.
\end{itemize} Then, the number $q$ of ions due to the $\gamma$ rays  of  Radio C in
air will be expressed as:
\begin{eqnarray*}
q & = & 2 \pi K Q \int_0^\infty \frac{r^2e^{-\lambda r}}{r^2} dr\\
q & = & 2 \pi \frac{K Q}{\lambda} = 0.035 \, .
\end{eqnarray*}

We should now  take into account the effect of the active products of Thorium, but there is no precise input
to complete the calculation. Eve assumes that, due to the effect
of the $\gamma$ radiation that is emitted by the products of 
Thorium, 0.025 ions per cm$^3$ per second are generated; which gives a   
total of 0.06 for the  ions in air.
In this calculation it is assumed that the air above the
sea surface has the same radioactive composition as the
air above the ground, 
while indeed at some distance from the coast, the content of radioactive emanation of the air above the sea is smaller with respect 
to the air above the soil, especially
regarding Thorium.

For the contribution from sea water, the calculation is
also easy, knowing from the Joly\footnote{Joly. Phyl. Mag., September 1909.} experiments that
the equivalent in Radio is $Q '= 1.1 \times 10^{-14}$; the absorption coefficient $\lambda'$
is  immediately obtained, using the relationship between
that coefficient  and the density:
\[ \frac{\lambda'}{\rho} = 0.034 \, \footnote{Mc. Clelland. Phil. Mag., July 1904.} \, . \]
Thus, we obtain the value $q = 0.006$ for  the sea.

To the value $q = 0.066$ we must add the effect of
secondary radiation emitted from the walls of the container;
we can assume that this will increase the effect that one has
in open air by 20\%, so we arrive at an estimate of a total ionization
that is on the order of 0.1 ions per cubic centimeter.

The measurements that I made on the sea had nonetheless
provided for values for $q$ that were, on average, significantly larger 
than those predicted by theory.
As an example, I take the
indications from the apparatus A \footnote{See Pacini, op. cit.}, which had walls 
that were 1.5 mm thick,
to exclude the vast majority of $\beta$ radiation.
Onboard
a boat with a surface of about 4 m$^2,$ this device provided an average measurement of
8.9 ions on the sea, with  a minimum of 4.7; in the hypothesis, supported by the results obtained so far, 
in which the minimum of 4.7 ions
can be ascribed entirely to the residual ionization, one is left with
an average of 4.2 ions, from which subtracting the action of
secondary radiation, we obtain the value:
\[ q = 3.4 \, \rm{ions} \]
 due to the penetrating radiation over the sea at a distance greater than 300
meters from the coast.

Subsequently, in May 1911,  
Simpson and Wright\footnote{G. C. Simpson and C. S. Wright:  \emph{Atmospheric Electricity over
the Ocean.} Proceed. of the Rovedo Soc. Vol. 85, p. 175, 1911.} published a note, reporting observations of
atmospheric electricity aboard the ``Terra Nova'' in a journey
from England to New Zealand, following the Antarctic expedition of
Captain Scott. 
The authors observed onboard
their ship, on average, a value of approximately 6 ions 
for the penetrating radiation; however, they found during several hours
values of approximately 9 ions after the ship
had left the coast -- an increase of 3 ions with respect to the
average value of $q.$ The minimum  value that was obtained for $q$ was
4 ions.

The results by Simpson and Wright confirm
that even outside of the direct action of soil it is possible to observe
considerable fluctuations in the values of penetrating radiation. The results 
of the experiments on which I will now report also seem to indicate the presence of well-measurable 
effects of 
penetrating radiation in air, over an
absorbing medium.

We shall see that by immersing the measuring device in
water, one can further lower the penetrating radiation that is observed at the surface
of a sea or lake below its average
value.

The A device, already used in the experiments above,
was enclosed in a copper box to be able to immerse it in water. The experiments were performed again
at the Naval Academy of Livorno, precisely
in the same place where the measurements of the previous year had been made. 

The apparatus was put onboard the same boat,
which was pegged at more than 300 meters from the coast, over 8-m-deep water.
Between June 24 and June 30, measurements were made
with the apparatus on the surface   and immersed at a depth of 3 meters.

Here are the results of these observations, each of which
had a duration of approximately 3 hours:

When  the instrument was at the surface, the loss per hour, measured in volts, was:
\[ 13.2 - 12.2 -  12.1 - 12.6 - 12.5 - 13.5 - 12.1 - 12.7\]
(average of 12.6, corresponding to 11 ions per cm$^3$ per second).

With the instrument immersed:
\[ 10.2 - 10.3 - 10.3 - 10.1 - 10.0 - 10.6 - 10.6 \]
(average of 10.3, corresponding to 8.9 ions per cm$^3$ per second).

The difference between these two values is 2.1 ions.

The boat was the same as the one that was used for the measurements in
which the minimum value of 4.7 ions was established, and because
 it had always been kept under the same conditions -- i.e., either on the
sea or suspended over the sea from the quay -- we are convinced that the boat did not contain
active materials other than  those that came from the air or 
sea. In the hours in which measurements were not made, the measuring apparatus
was kept charged, always in the same room, and the dispersion
of electricity was strictly constant.

With the same apparatus, measurements were also made at the
Lake of Bracciano. At 350 meters from the shore,  I measured 
$q = 12.4 q$ at the surface, while at a depth of 3 meters
(at a place where the bed 
was over 7 m deeper), the result was
$q = 10.2.$ Thus, the difference in the two values of $q$ was 
2.2 ions.

With an absorption coefficient of 0.034  for water, it is 
easy to deduce from the known equation $I/I_0 = e^{-\lambda d}$, where $d$ is the
thickness of the matter crossed, that, in the conditions of my
experiments, the activities of the bed and of the surface were both
negligible.

The water temperature was, on average, a few tenths of a degree
lower than the air above, and working under
airtight conditions, the number of ions created in the internal space
 varies only due to changes in radiation. From the measured  differences of
2.1 and 2.2, by subtracting 20\% due to secondary radiation,
these numbers become:
\begin{center}
\begin{tabular}{l}
1.7 ions for the sea\\
1.8 ions for the Lake of Bracciano.
\end{tabular}
\end{center}

Is this decrease in the value of $q,$ moving from observation at
the surface to a survey of the interior of the water,
 due to external actions or rather to a variation in the residual
ionization of the container in the transition from air to water?

We know nothing for certain about the origin of the
residual ionization for air  that is trapped in a metallic container.

The causes that might generate the
residual ionization  
are intrinsic activity, or radioactive impurities of the
metal, and possible spontaneous ionization  of the enclosed gas\footnote{G.C. Simpson and C.S. Wright (op. cit.).}.

 Under the conditions in which 
these experiments are conducted, it is unlikely that metals, with the exception of
lead, contain radioactive impurities. Furthermore, in a
long series of observations that were made earlier with the same apparatus, an increase in dispersion
that could be ascribed to impurities was never observed.

In the case of activity that is due to the metal or 
spontaneous emission of electrons due to disintegration of the gas
that is encapsulated in the device, one cannot see any reason for a variation
of these causes of ionization, given the changing
conditions of  the isolated instrument between the surface
and the depth.

The explanation appears to be that
 due to the absorbing power  of water and the minimum amount
of radioactive substances in the sea, absorption of $\gamma$ radiation
coming from the outside happens indeed, when the apparatus is immersed.

It is natural, as already pointed out\footnote{D. Pacini (op. cit.).}, 
to look for the origin of such ionization of  air due to penetrating radiation, not directly dependent on 
active substances in the soil, 
in an accumulation of radioactive material that released into the atmosphere around the
site  of observation.

Simpson and Wright also attribute
an increase of three ions to this cause, in their measurement of normal ionization on the sea.
According to these authors, the active particles could have
deposited onto the ship from the air when the ship was near the coast.

If we assume that the products are spread evenly in the atmosphere at altitudes of up to 5 km, and that they
are rapidly deposited on the surface
of the Earth from the air,  from Eve's  data one deduces a 
layer of Ra. C that is equivalent to  $4 \times 10^{- 11}$ of
Radium per cm$^3$ in equilibrium.  This would generate 1.8 ions per cm$^3$ per second in the air at a height of 1 meter.

In the case of my experiments, one can neglect the action of
active particles that deposit onto the water, because they should go into solution quickly, due to  wave motion.
We can get a sense of what the effect would be of an
active substance that is deposited onto the boat used here, that has a surface of approximately 4
m$^2.$ Suppose that the Ra. C that is deposited onto the
boat acts on the device (which is located in the center, above
a table, at the height of the edge) as if it were distributed evenly
on the surface of a half-sphere whose radius is 80 cm, in an amount $Q$ that is equivalent to 
$4 \times 10^{-11}$ grams of Ra per cm$^2$. The number of ions that is generated by the full radioactive
deposit in 1 cm$^3$  of air in the
center of the hemisphere would be expressed as
\[ q  =  \frac{K Q}{r^2} e^{-\lambda r}2 \pi r^2= 0.8 \, \rm{ions} \, . \]
Assuming that the products of Thorium are responsible for 0.5 ions in this
case, we would have a total of 1.3 ions. This calculation gives us
a value smaller than that observed, but the effect is nevertheless
well measurable.

A rapid reduction in the active products of the atmosphere
could occur for large values of the Earth's field, especially
in the case of rain. The observations that have been made so far
on the behavior of penetrating radiation during
rain are not quite in agreement, and they are not enough
to establish the existence of an
action in the sense mentioned above.

Free-balloon experiments have been performed recently\footnote{A. Goekel. Phys. Zeit., p. 595, 1911 and V.F. Hess. Phys. Zeit., p. 998, 1911.} 
on penetrating radiation in the upper atmosphere. 
Although they cannot be considered
conclusive with regard  to the study of the radiation that penetrates
at a certain height above the ground, these observations, however, could have
shown that where, according to the law of absorption from air
(recently verified by Hess), the action of
active substances of the soil is negligible, there is still a large quantity of penetrating radiation. This result has
spurred Gockel and Hess to repeat what the author of the present paper concluded from the first observations that were made on the sea and
what appears from the results of the work described in this Note:
that \emph{a sizable cause of ionization exists in the atmosphere, originating from penetrating radiation, 
independent of the 
direct action of radioactive substances in the soil.}

\vspace{1cm}\noindent

\noindent{\small{[Note by A. De Angelis: Thanks to Dr. Stefania De Angelis from Williams Language Solutions and Dr.
Sean Kim from Blue Pencil Science for help in the translation and
editing; 
and to the collegues N.~Giglietto, S.~Stramaglia, A.~Garuccio, L.~Guerriero, E.~Menichetti, P.~Spinelli, F.~Guerra, N.~Robotti, L.~Cifarelli and P.~Carlson for discussions and for material about the work of Pacini.]}}

\newpage

\section*{Reprint of the original article, with the kind permission of the President of the Societ\`a Italiana di Fisica, professor Luisa Cifarelli.} 


\hspace{-0.7cm} \includegraphics{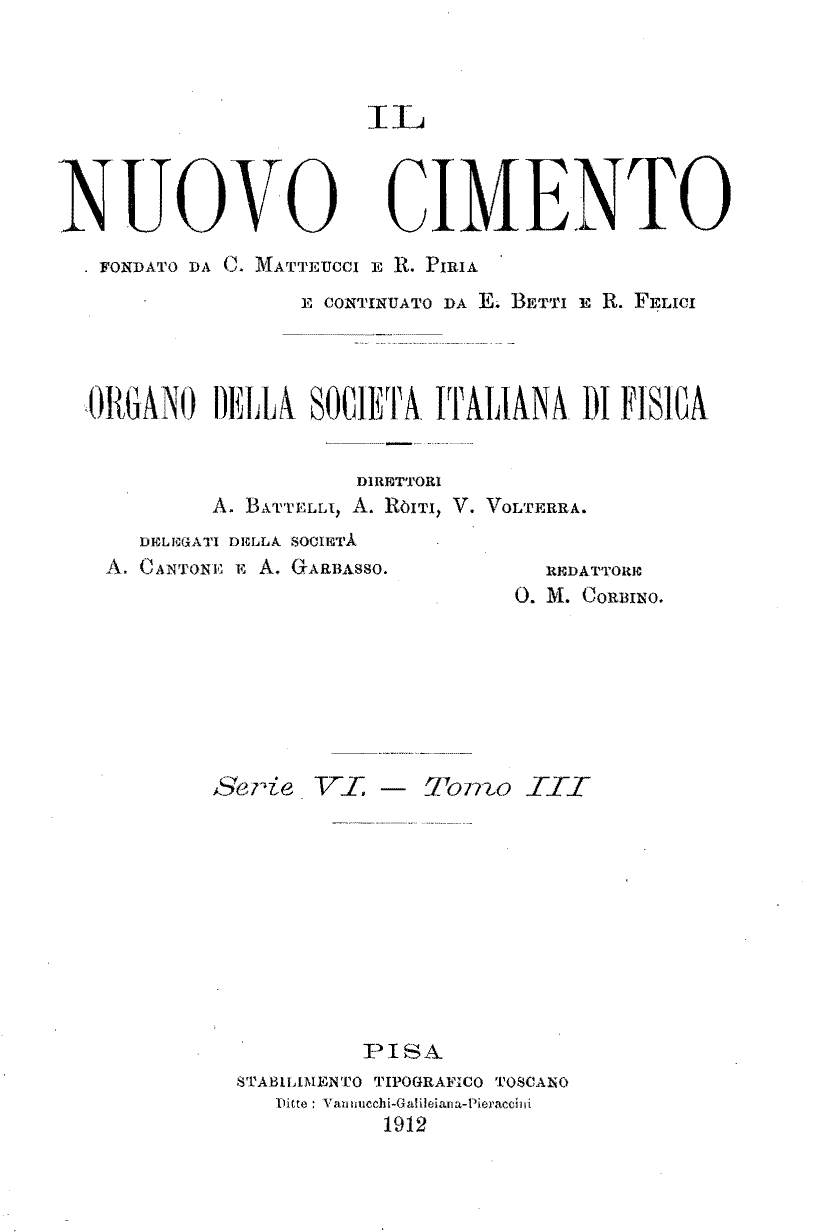} \newpage
\hspace{-0.7cm}\includegraphics{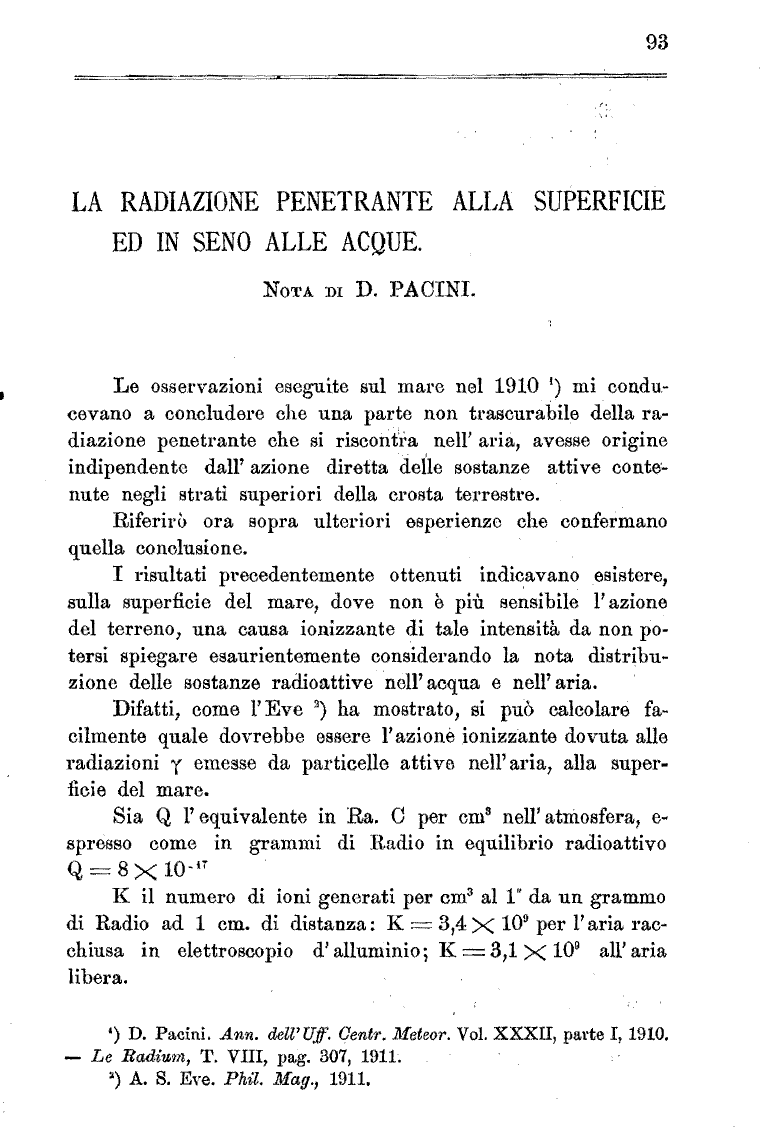} \newpage
\hspace{-0.7cm}\includegraphics{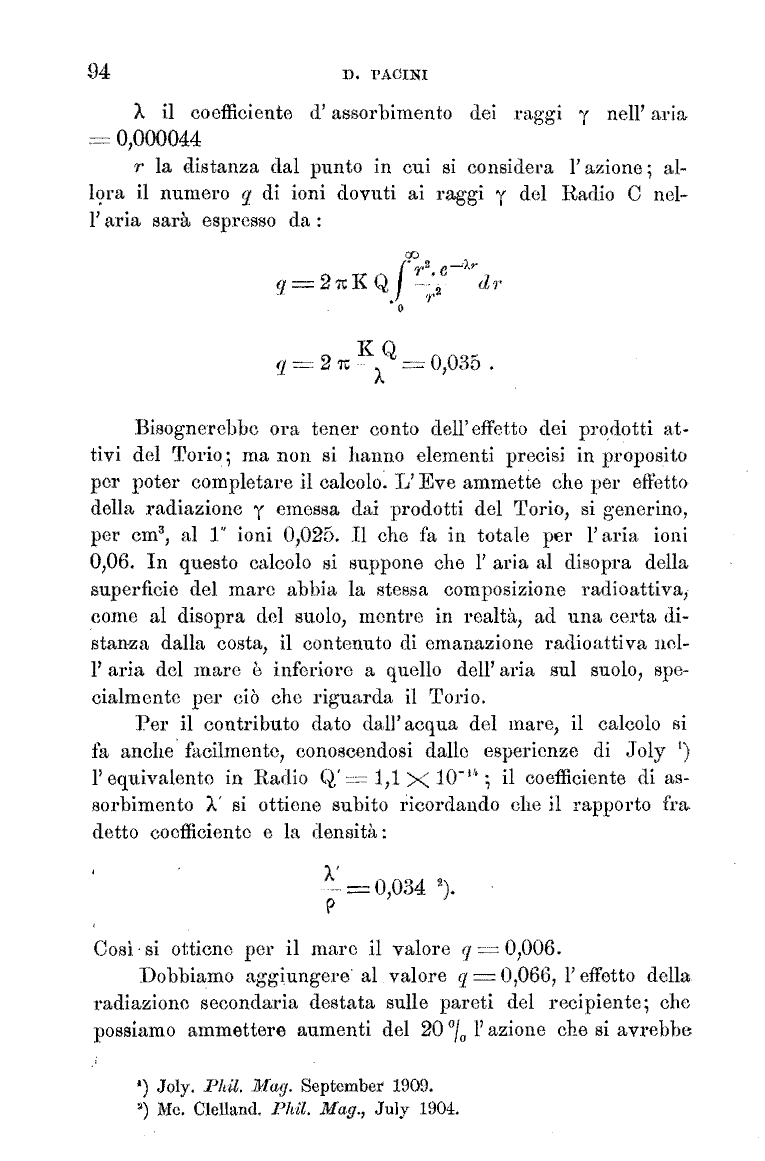}\newpage
\hspace{-0.7cm}\includegraphics{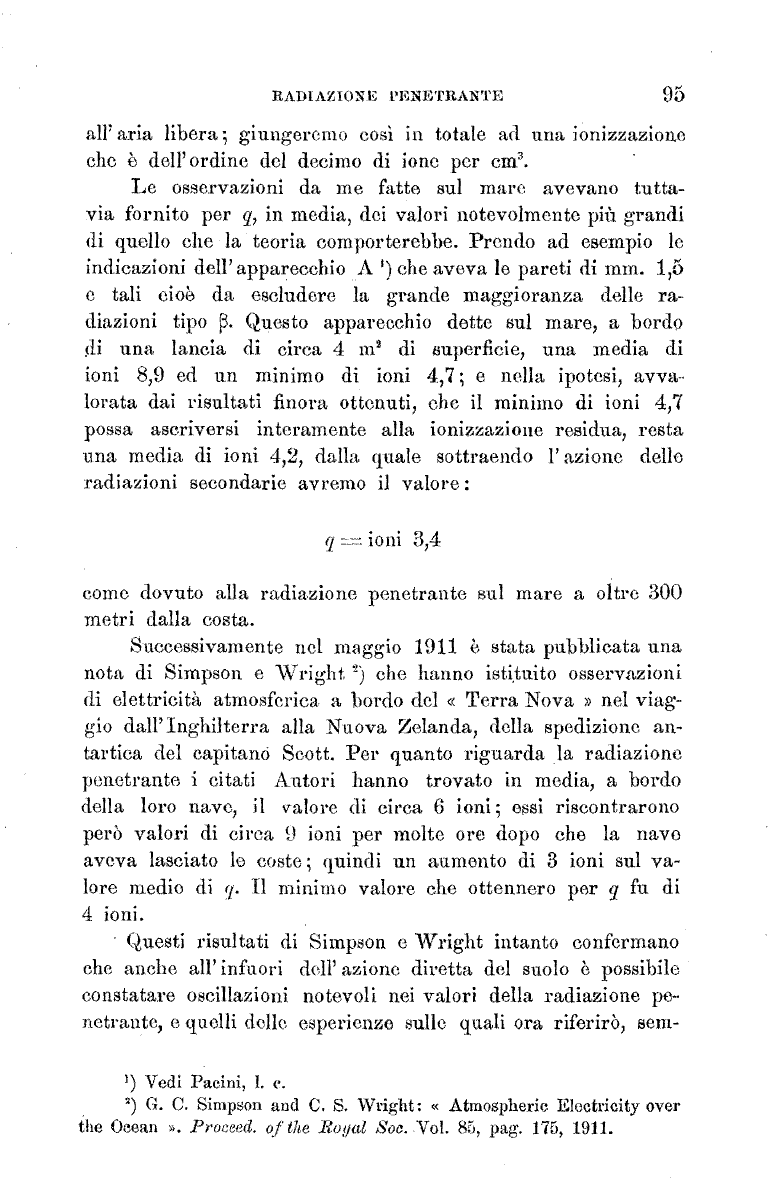}\newpage
\hspace{-0.7cm}\includegraphics{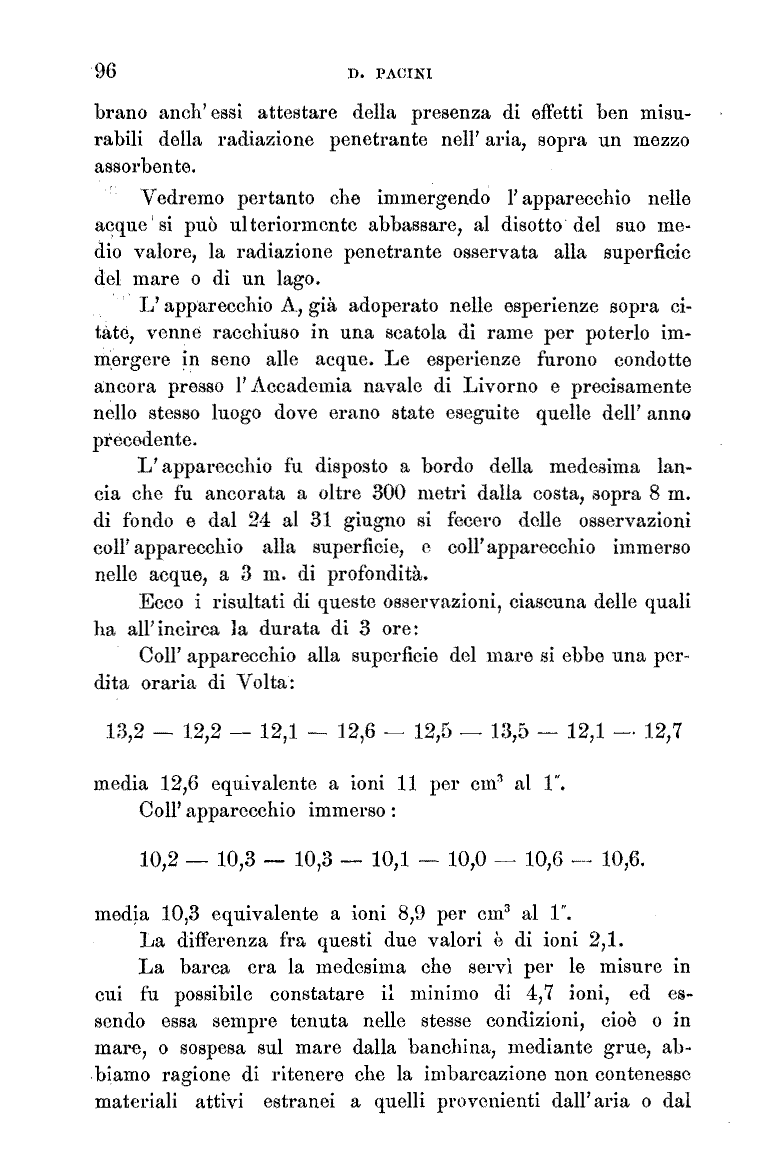}\newpage
\hspace{-0.7cm}\includegraphics{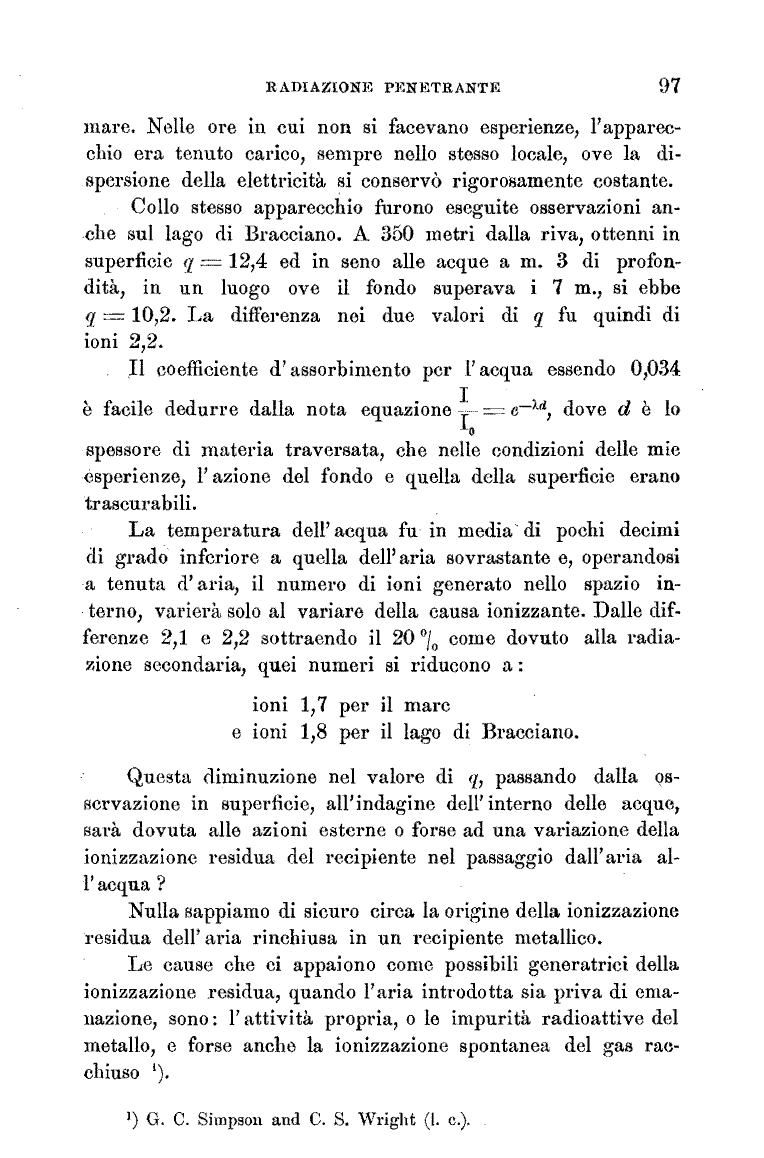}\newpage
\hspace{-0.7cm}\includegraphics{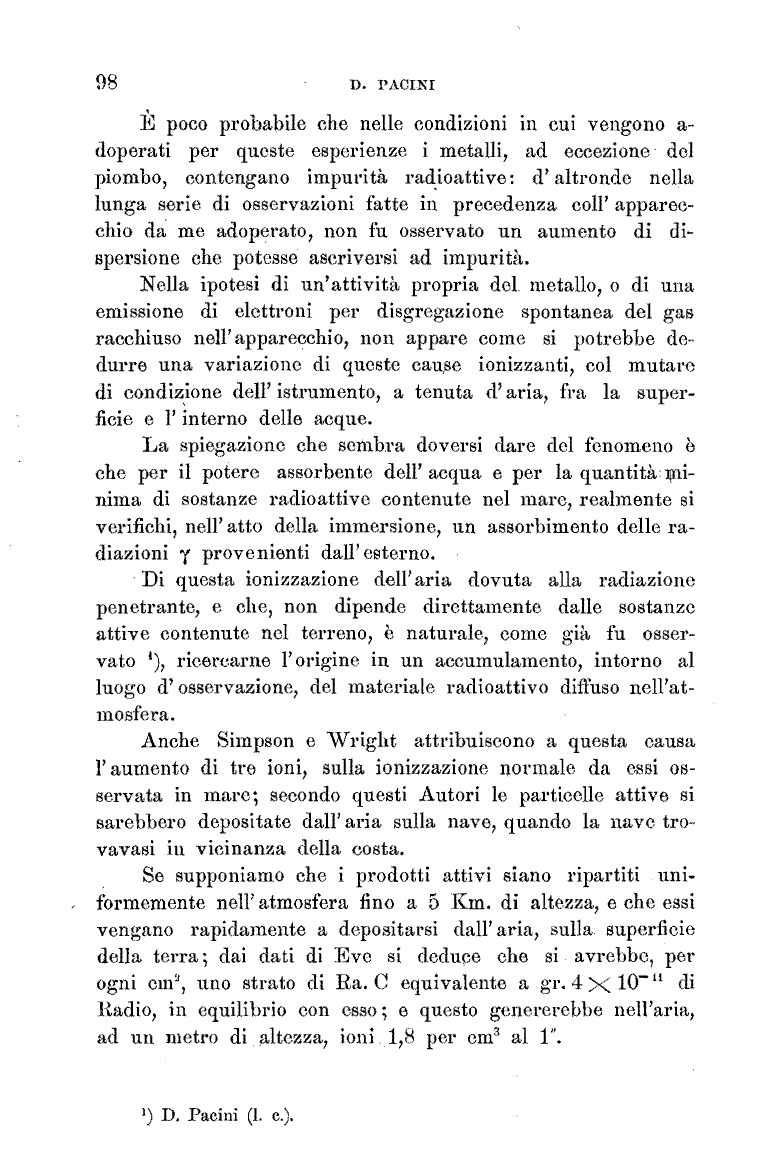}\newpage
\hspace{-0.7cm}\includegraphics{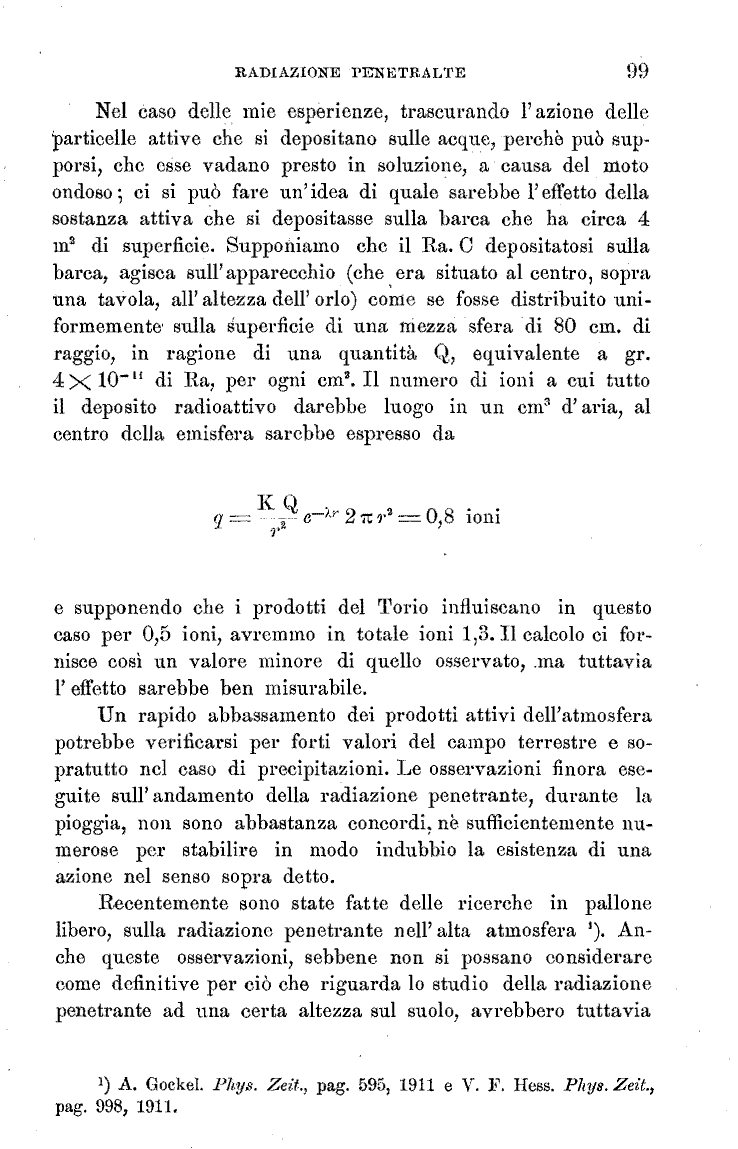} \newpage
\hspace{-0.7cm}\includegraphics{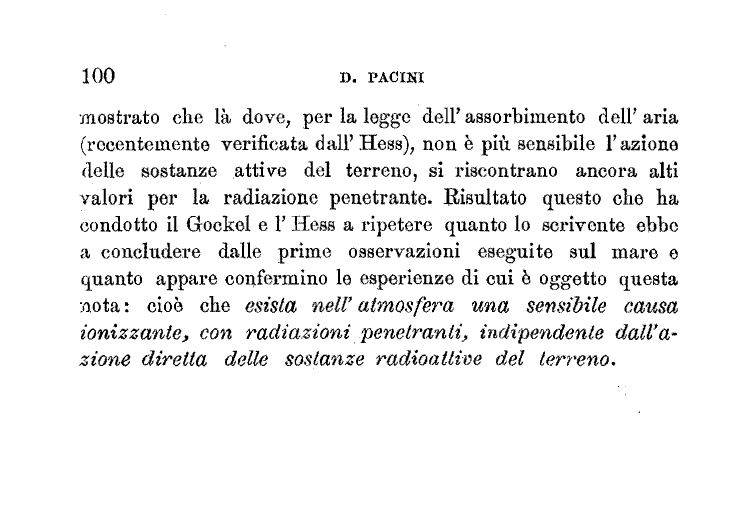}

\end{document}